\documentclass[pre,superscriptaddress,tightenlines,showpacs,nofootinbib,notitlepage,twocolumn]{revtex4-2}
\usepackage{graphicx,amssymb,amsmath,epsf,bm}
\usepackage[normalem]{ulem}
\usepackage{stackrel}
\usepackage{color}
\usepackage{upgreek}

\definecolor{nblue}  {RGB}{28,130,185}

\definecolor{cgreen}  {RGB}{76,153,0}

\usepackage{latexsym,amsmath,amsfonts,tabularx,amssymb,bm,empheq}
\usepackage{graphicx,epstopdf}
\usepackage{xspace}

\usepackage{hyperref}

\newif\ifhyper
\hypertrue
\ifhyper
\hypersetup{
  citecolor = {green},
  colorlinks = {true}, 
  urlcolor = {blue} 
}

\fi

\newcommand{\bea}{\begin{eqnarray}}
\newcommand{\eea}{\end{eqnarray}}

\def\K{\mathbf{K}}
\def\bs{\boldsymbol{\sigma}} 
\def\bmu{\boldsymbol{\mu}}

\def\RR{\mathcal{R}}

\def\simge{\mathrel{%
   \rlap{\raise 0.511ex \hbox{$ \rangle $}}{\lower 0.511ex \hbox{$\sim$}}}}
\def\simle{\mathrel{
   \rlap{\raise 0.511ex \hbox{$ \langle $}}{\lower 0.511ex \hbox{$\sim$}}}}

\def\simle{\mathrel{
   \rlap{\raise 0.511ex \hbox{$ \langle $}}{\lower 0.511ex \hbox{$\sim$}}}}

\def\simge{\mathrel{%
    \rlap{\raise 0.511ex \hbox{$ \rangle $}}{\lower 0.511ex \hbox{$\sim$}}}}

\usepackage{subfigure}
\usepackage{color}
\usepackage[dvipsnames]{xcolor}
\usepackage{soul}
\usepackage{tabularx}

\begin{document}

\title{Dreaming up scale invariance via inverse renormalization group}

\author{Adam Ran\c con}
\email{adam.rancon@univ-lille.fr}
 \affiliation{
Univ.\ Lille, CNRS, UMR 8523 -- PhLAM -- Laboratoire de
Physique des Lasers Atomes et Mol\'ecules, F-59000 Lille, France
}
\affiliation{Institut Universitaire de France}
\author{Ulysse Ran\c con}%
 \email{ulysse.rancon@cnrs.fr}
\affiliation{CerCo UMR 5549, CNRS – Université Toulouse III, Toulouse, France}%


\author{Tomislav Ivek}
\email{tivek@ifs.hr}
\affiliation{Institute of Physics, Bijeni\v{c}ka cesta 46, HR-10001 Zagreb, Croatia}

\author{Ivan Balog}
\email{balog@ifs.hr}
\affiliation{Institute of Physics, Bijeni\v{c}ka cesta 46, HR-10001 Zagreb, Croatia}


\begin{abstract}
We explore how minimal neural networks can invert the renormalization group (RG) coarse-graining procedure in the two-dimensional Ising model, effectively ``dreaming up'' microscopic configurations from coarse-grained states. This task—formally impossible at the level of configurations—can be approached probabilistically, allowing machine learning models to reconstruct scale-invariant distributions without relying on microscopic input. We demonstrate that even neural networks with as few as three trainable parameters can learn to generate critical configurations,  reproducing the scaling behavior of observables such as magnetic susceptibility, heat capacity, and Binder ratios. A real-space renormalization group analysis of the generated configurations confirms that the models capture not only scale invariance but also reproduce nontrivial eigenvalues of the RG transformation. While the inversion is necessarily imperfect, these minimal models robustly reproduce the RG-relevant structure of the critical distribution. Surprisingly, we find that increasing network complexity by introducing multiple layers offers no significant benefit. These findings suggest that simple local rules, akin to those generating fractal structures, are sufficient to encode the universality of critical phenomena, creating an opportunity for efficient generative models of statistical ensembles in physics.  
\end{abstract}

\maketitle

\section{Introduction}

Universal behavior near critical points is a cornerstone of statistical physics. In systems undergoing second-order phase transitions, observables such as correlation functions and susceptibilities exhibit power-law scaling that is largely independent of microscopic details. The renormalization group (RG) framework elegantly explains this universality by describing how short-range fluctuations are systematically averaged out to reveal long-range scale-invariant properties \cite{Wilson1974}. Yet, this very process of coarse-graining is inherently lossy and one-way: 
it filters out short-distance details while reorganizing the remaining information into emergent degrees of freedom at larger scales and cannot, in general, be inverted. However, it bears a striking resemblance to various techniques in Deep Learning (DL) that also focus on compressing the degrees of freedom and keeping the most relevant ones. This connection has been under intensive scrutiny recently \cite{Beny2013,Mehta2014,MelloKoch2020,Gordon2021,Gokmen2021,Erdmenger2022,Kline2022,Marchand2023,biroliDynamicalRegimesDiffusion2024,masuki_generative_2025}.

Machine learning and DL have become foundational tools across modern science and engineering. These methods have dramatically improved the state-of-the-art in tasks ranging from image and speech recognition to drug discovery and genomics, and they have entered many scientific disciplines as indispensable techniques for data analysis and modeling \cite{Carleo2019}.
Over the last decade, deep learning has emerged as a powerful tool for data transformation, particularly in the domain of image processing where super-resolution plays a central role in reconstructing fine-scale structures from coarse data \cite{pixelshuffle, yang2019deep}. Models based on convolutional networks have demonstrated remarkable success in upscaling complex data structures, both in general imaging applications and in contexts where outputs must respect physical constraints \cite{fukami2021machine,SRcosmo}. In particular, generative models such as generative adversarial networks \cite{NIPS2014_f033ed80}, variational autoencoders \cite{kingma2022autoencodingvariationalbayes}, and, more recently, diffusion models \cite{NEURIPS2020_4c5bcfec}, have significantly advanced the ability to synthesize high-resolution, high-dimensional data from low-dimensional inputs by learning the underlying data distributions \cite{ledig2017photo,esser2021taming,rombach2022high}. These models enable the inference of plausible fine-grained details, yielding high-quality reconstructions.

\begin{figure}
    \centering
    \includegraphics[width=1.0\linewidth]{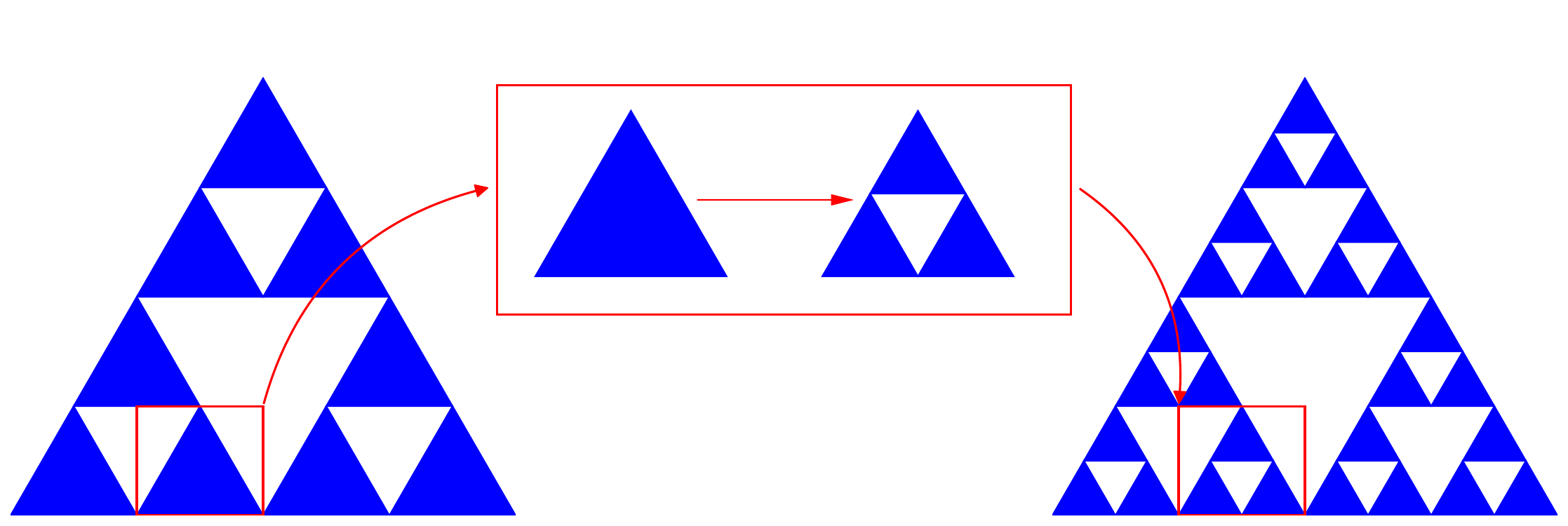}
    \caption{Sierpi\'nski triangle is a self-similar structure (a fractal) generated by an iterative application of a simple local rule (inset). A critical configuration of spins constitutes a fractal object as well, which leads us to explore the concept of using simple local neural networks for dreaming up critical spin configurations.}
    \label{fig:local_rule_fractal}
\end{figure}

This capability is naturally aligned with the challenge of inverting RG transformations, which has been the focus of recent works \cite{Efthymiou2019,Bachtis2022,Shiina2021,Bachtis2024a,Bachtis2024b}, demonstrating that complex neural networks trained on critical small system configurations can generate configurations for larger systems that demonstrate scale invariance but also obtain reasonable critical exponents. At the same time, phase-characterization models have evolved from early supervised and unsupervised architectures, using techniques like principal component analysis and autoencoders to detect transitions and learn emergent order parameters directly from raw Ising spin data \cite{Wang2016,Wetzel2017}, to increasingly complex structures such as graph neural networks that encode interaction topology \cite{Ma2024}, attention‐based transformers that capture long-range correlations \cite{Kara2021}, and symmetry-equivariant architectures enforcing lattice and spin-flip invariances \cite{Chang2023}. While such approaches often greatly increase model complexity, it remains unclear if they are useful for inverting RG transformations or indeed what the minimal conditions are for a neural network to learn to generate configurations that are scale invariant. A related line of work, notably that of Marchand et al.~\cite{Marchand2023}, addresses a different inference problem: Given microscopic spin configurations,  the goal is to reconstruct the underlying Hamiltonian or probability distribution that generated the data. Their framework employs a hierarchical,  RG-inspired decomposition of the fine-scale distribution into a set of local,  conditional models. 
In this sense, their approach is broader in scope, as it aims to recover the full hierarchy of conditional probabilities at all levels of the wavelet decomposition. By contrast, the present work focuses on the intrinsically ill-posed task of probabilistically inverting real-space coarse graining: Our objective is not to recover the microscopic Hamiltonian, but to determine what machine learning models learn when explicitly trained to generate plausible microscopic configurations from coarser ones over and over again.

In this work we revisit this problem from a minimalistic perspective. Rather than employing deep or highly parametrized models as in \cite{Efthymiou2019,Bachtis2022,Shiina2021,Bachtis2024a,Bachtis2024b,Bauer2025}, we ask the following question: What is the simplest neural network that can learn to invert RG and reproduce scale-invariant physics? By training single-layer convolutional networks with as few as three parameters, we show that it is possible to generate critical Ising configurations that reproduce key thermodynamic observables and universal finite-size scaling relations. Remarkably, we find that such simple models can match and in some respects outperform more complex architectures. The resulting configurations obey universal scaling laws and display stable RG eigenvalue spectra, confirming that even minimal models can learn the essential features of criticality.
That a very simple network, with straightforward transformation rules, is able to produce scale invariance might not be too surprising in hindsight. Indeed, this is reminiscent of constructing fractals by local rules, such as the Sierpi\'nski triangle, where simple iterative procedures yield complex, self-similar structures such as that illustrated in Fig.~\ref{fig:local_rule_fractal}. 

Here, we focus on the Ising model, the standard model of statistical physics, known for its simplicity and rich critical behavior that makes it a classic example of universality in phase transitions. The model describes a two-dimensional lattice of spins, which interact via a Hamiltonian
\begin{equation}
    H(\bs)=-K\sum_{\langle i,j\rangle} \sigma_i \sigma_j,
\label{eq:Ising}
\end{equation}
with ferromagnetic interaction $K$ (proportional to the inverse temperature), where the sum is over nearest neighbors. This model undergoes a phase transition at a critical coupling,  $K_c$, separating the paramagnetic phase, where spins are weakly correlated, from the ferromagnetic phase, where spins align. Furthermore, the physics close to the transition exhibits a universal behavior characterized by divergences in quantities such as magnetic susceptibility and heat capacity. These quantities follow power-law scaling with critical exponents that are independent of the system’s microscopic details—a hallmark of universality. Despite its simplicity, the Ising model’s critical properties are applicable to a broad class of physical systems, making it a paradigmatic example of a universality class in statistical physics. 

Our objective in this study is to determine whether simple neural networks, with minimal parameters, can capture the scale-invariant properties of the Ising model at criticality. While such simple neural nets will never reproduce the exact sampling of spin configurations, we explore whether such networks can dream up configurations that reflect the universal characteristics of critical systems. Inspired by fractal patterns, which arise from simple iterative rules, we focus on a minimalistic network architecture with linear filters of varying kernel sizes, assessing their capacity to learn universal scaling.

To validate our approach, we perform a series of tests examining how well these networks reproduce known critical behavior. We start by calculating the scaling properties of standard quantities such as order parameter, susceptibility, heat capacity etc., but also examine the critical distribution of the order parameter. For a more stringent analysis, we conduct a real space renormalization group (RSRG) analysis on upscaled configurations to test the networks' learning of scale invariance. Our surprising results show that even these simple networks can approximate universal features of criticality effectively, often outperforming more complex models.

The paper is organized as follows. Section \ref{sec:upscaling} describes the concepts of downscaling and upscaling and their connection to the renormalization group and scale invariance. In Sec.~\ref{sec:model}, we introduce the simple neural network we study, its training and the upscaling procedure. In Sec.~\ref{sec:FSS}, we demonstrate the ability of this net to capture scale invariance by studying finite-size scaling of various observables, while in Sec.~\ref{sec:MCRG} we show that it also captures the relevant properties of the renormalization group. In Sec.~\ref{sec:complicated} we discuss how more complex models do not perform better to these tasks. We provide a summary and discuss our results in Sec.~\ref{sec:conclusion}.

The code used for model training and evaluation is publicly available in Ref.~\cite{Rancon_upscaling_Ising}.

\section{\label{sec:upscaling} Downscaling, upscaling, and scale invariance}

The probability distribution of a generic Ising model of linear size $L$ for a given spin configuration $\bs=\{\sigma_i\}_{i=1,\ldots,L^2}$, $\sigma_i=\pm1$, can be written as
\begin{equation}
    P(\bs) = e^{-H(\bs)},
\end{equation}
in terms of the generic Hamiltonian (the temperature is factored in the Hamiltonian)
\begin{equation}
    H(\bs) = \sum_\alpha K_{\alpha} S_{\alpha}(\bs).
\end{equation}
Here $S_{\alpha}$ are local spin operators (for instance nearest neighbor interaction) and $K_{\alpha}$ are the corresponding coupling constants. The normalization of $P$ is given by $K_{0}$ (with $S_{0}(\bs)=1$).

Kadanoff's blocking procedure from a spin configuration $\bs$ to a (coarse-grained) block-spin configuration $\bmu$ can be reframed in a probabilistic setting by introducing the coarse-graining kernel (the conditional probability of a block spin) $D(\bmu|\bs)$ and a joint probability $P(\bs,\bmu) =  D(\bmu|\bs)P(\bs)$. Assuming the block-spins are constructed such that the linear size of the system is halved, one constructs the probability distribution of the block-spins as
\begin{equation}
    P_{1}(\bmu) = \sum_{\bs} P(\bs,\bmu)= \sum_{\bs} D(\bmu|\bs) P(\bs).
\end{equation}
The coarse-graining kernel $D$ is chosen such that $\sum_{\bmu} D(\bmu|\bs)=1$ satisfies the so-called trace condition, $\sum_{\bmu}P(\bs,\bmu)=P(\bs)$, to leave the thermodynamics of the system invariant. This transformation can be deterministic or probabilistic but is not invertible: it should allow for forgetting irrelevant (microscopic) information.
In the following, we will consider the standard majority rule to block-spin, i.e.\ the lattice is divided into $2\times2$ blocks where each block is transformed to a spin with the same state as the majority of spins in the block. In case of ambiguity, we randomly pick the block-spin to be $\pm 1$ with probability $1/2$. The blocking procedure can also be interpreted in probabilistic and information theoretical setups in terms of restricted Boltzmann machine or more complicated neural nets and optimized upon \cite{Koch-Janusz2018,Lenggenhager2020,Chung2021,Gokmen2021,Gordon2021,Wu2017}.

The block-spin probability distribution $P_{1}(\bmu)$ can be written in terms of a block-spin Hamiltonian
\begin{equation}
    H_{1}(\bmu) = \sum_\alpha K_{\alpha,1} S_{\alpha}(\bmu).
\end{equation}
Here we assume that the spin operators $S_{\alpha}$ do not depend on the system size (i.e.\ they are all local and do not involve interactions between spins that are at a distance of the order of the system size), which is fine as long as the system size $L$ is very large.

By performing $n$ blocking procedure, one generates a flow of coupling constants $\K=\{K_{\alpha}\}$ given by the trajectory of $K_{\alpha,0}\to K_{\alpha,1}\to\ldots\to K_{\alpha,n}$, with $K_{\alpha,0}\equiv K_\alpha$ the microscopic couplings. This implements an RSRG transformation on the couplings $\K_{n+1}=\RR(\K)$. Critical points of second-order phase transitions and scale-invariance are found by looking for the fixed point of the blocking procedure $K_{\alpha,\star}$, such that $\lim_{n\to\infty}K_{\alpha,n}= K_{\alpha,\star}$.  Equivalently, we then have $P_n\to P_\star$, with
\begin{equation}
    P_{\star}(\bmu) = \sum_{\bs} D(\bmu|\bs) P_\star(\bs).
\end{equation}
This is the hallmark of scale invariance: After sufficiently many coarse-graining steps, $P_n$ flows to $P_\star$, the fixed point of the coarse-graining procedure.

Obviously, the fixed point equation can only make sense if the system size is strictly infinite (such that the number of spins does not change after the blocking procedure). For very large but finite systems, $P_n$ is expected to approach the fixed point probability distribution for $n\gg 1$ and $L/2^{n}\gg1$, while finite size effects (and departure from the fixed point) will be generated when $L/2^{n}\simeq 1$.

The goal of the upscaling algorithm is to learn how to undo the blocking (i.e.\ downscaling) procedure. This is done by introducing another conditional probability $U(\bs|\bmu)$ which attempts to undo the effect of $D(\bmu|\bs)$. Note that we do not require that it undoes it exactly, i.e.\ $\sum_{\mu}U(\bs|\bmu)D(\bmu|\bs')=\delta_{\bs,\bs'}$, nor do we aim for it, as it is impossible by construction since the information about the specific spin configuration in a block is lost. Instead, this should be (approximately) true at the distribution level, i.e.\ we aim for the best $U(\bs|\bmu)$ such that
\begin{equation}
    Q(\bs,\bmu)= U(\bs|\bmu) \sum_{\bs'} D(\bmu|\bs')P(\bs')
    \label{eq:Q}
\end{equation}
is a good approximation of the true joint probability $P(\bs,\bmu)$, i.e.\ $Q(\bs,\bmu)\simeq P(\bs,\bmu)$. From Eq.\ \eqref{eq:Q}, we find that given a block-spin configuration $\bmu$, with probability $P_1(\bmu) = \sum_{\bs}P(\bs,\bmu) =\sum_{\bs} D(\bmu|\bs)P(\bs)$, we estimate the probability distribution of the underlying microscopic distribution to be $Q(\bs)=\sum_{\bmu }Q(\bs,\bmu)=\sum_{\bmu }U(\bs|\bmu)P_1(\bmu)$. In other words, we are able to generate (``dream up'') fine-grained configurations from coarse-grained ones.

Note that while the coarse-graining kernel $D$ is independent of the (microscopic) couplings $K_\alpha$, and thus of the temperature, it cannot be so for the upscaling kernel $U$. Indeed, for a given block spin, the typical underlying configurations of the (true) spins will be strongly temperature-dependent (quite random at high temperatures, very polarized at low temperatures), and this has to be learned during training. We focus here on criticality to test how well a system trained that way can generate scale-invariant configurations. This is indeed in principle possible since, if the kernel is general enough and properly trained, we should find that
\begin{equation}
\label{eq:Pstar}
    P_\star(\bs) \simeq  \sum_{\bmu} U(\bs|\bmu) P_\star(\bmu).
\end{equation}
As we will see, already very simple kernels, as small as having three parameters, can achieve this with high fidelity. Furthermore, we will show that sampling from Eq. \eqref{eq:Pstar} generates faithful critical configurations without any input from Monte Carlo simulations (beyond training).

\section{\label{sec:model} Models, training and upscaling}

\subsection{Models}

Machine learning and neural networks are tailored for optimizing such problems and have been used recently in this context \cite{Efthymiou2019,Shiina2021,Bachtis2022}. However, the complicated multilayer neural nets used there prevent a simple understanding of what has been learned. Here we focus instead on a very simple, one-layer convolutional network that already captures very well scale invariance and the renormalization group properties of coarse-graining.

\begin{figure}
    \centering
    \includegraphics[width=1\linewidth]{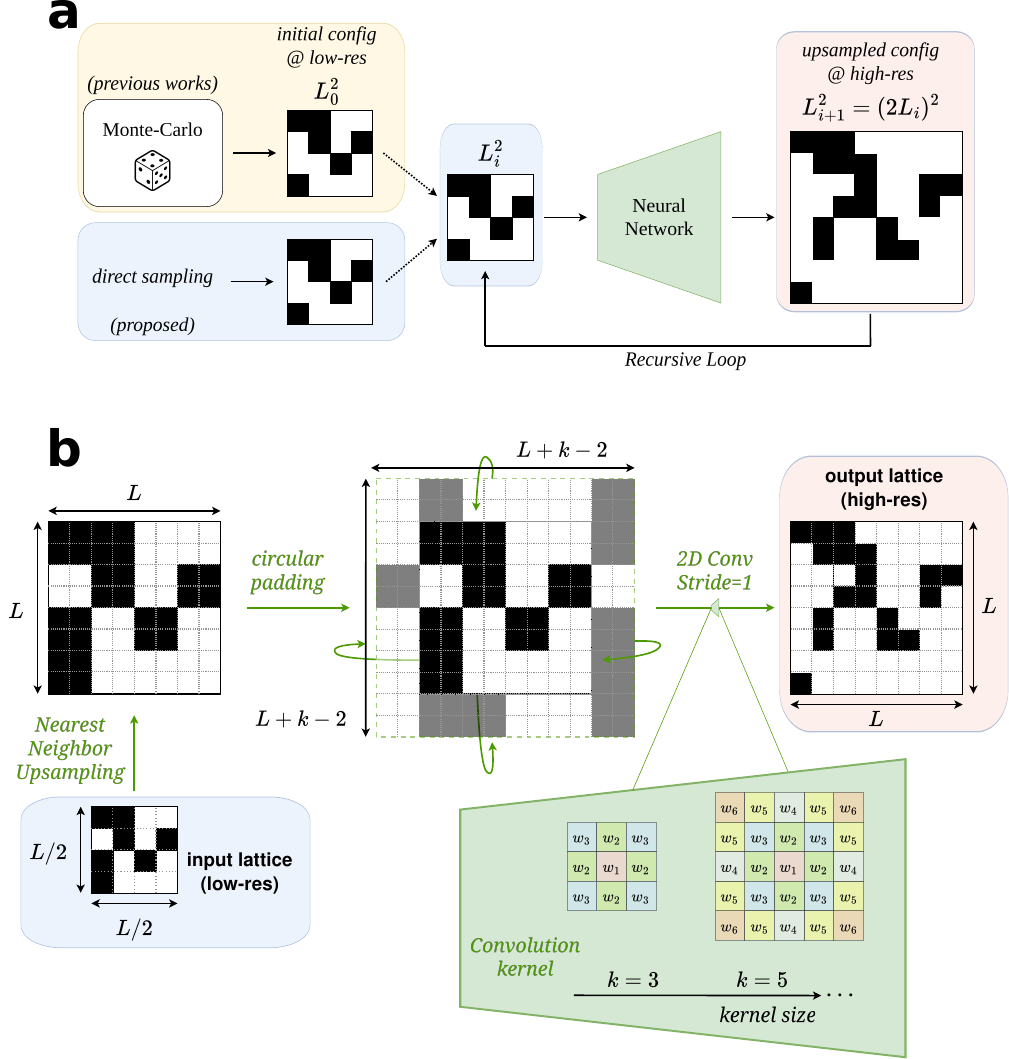}
    \caption{Iterative upscaling procedure and proposed convolutional network architecture. (a) A trained network maps a coarse configuration of size $L_i \times L_i$ to one of size $L_{i+1}\times L_{i+1}=4L_i\times L_i$. Previous works (top row) start from Monte-Carlo-sampled configurations at criticality. In our scheme (bottom row), the same network can be iteratively applied to a single block spin $L_0=1$ so that large microscopic configurations are generated solely by repeated applications of the conditional upscaling kernel $U(\bs|\bmu)$. (b) One upscaling step from an $L/2 \times L/2$ block-spin configuration $\bmu$ (left) to an $L \times L$ spin configuration $\bs$ (right). Nearest-neighbor upsampling first duplicates each spin into a $2 \times 2$ block and produces an intermediate field $\tilde{\bs}$ (middle panel). The light gray shaded spins outside the red dashed square are periodic copies of the boundary spins of $\tilde{\bs}$. Together with the curved arrows, they indicate the periodic boundary conditions respected by the convolution (circular padding). When the kernel is centered close to an edge, it wraps around and samples spins from the opposite side of the lattice. A single 2D convolution with stride one and circular padding (light green block) then computes the local fields entering the logistic conditional probabilities $p(\sigma_i|\bmu)$, from which the final spins $\sigma_i$ of the output configuration are drawn independently. 
    }
    \label{fig_model}
\end{figure}

We choose a simple convolutional layer with a kernel that upscales a given $L/2\times L/2$ configuration $\bmu$ as follows, see Fig.\ \ref{fig_model}. We first upscale the configuration to a $L\times L$ configuration $\tilde\bs$ by transforming each up (down) spin to a block of four up (down) spins (this rule is known as the ``nearest-neighbor'' rule in machine learning). This can be implemented using a kernel $W_{nn}$, with $\tilde\bs = W_{nn}\bmu$ ($W_{nn}$ is a $L^2\times (L/2)^2$ matrix). While this configuration $\tilde\bs$ could be a good guess at low temperatures (where spins in the same block are expected to point in the same direction), this is not so at higher temperatures (and in particular at criticality). To create correlations between spins belonging to different blocks, we use a convolutional network parametrized by a convolution kernel $W_c$ (a $L^2\times L^2$ matrix in general). Defining $W=W_cW_{nn}$ (matrix multiplication implied), the probability distribution of the spin $\sigma_i$ given $\bmu$ is then given by
\begin{equation}
\label{eq:Psm}
    p(\sigma_i|\bmu)=\frac{e^{\sigma_i (W\bmu)_i}}{\sum_{\sigma_i}e^{\sigma_i (W\bmu)_i}}.
\end{equation}
Note that if $\sigma_i$ were a binary variable ($0$ or $1$), this probability density would then be described by a sigmoid function, i.e. it is a logistic model.
The conditional probability to upscale is thus given by $U(\bs|\bmu)=\prod_i p(\sigma_i|\bmu)$.

We do not include a bias in Eq.\ \eqref{eq:Psm} (i.e linear term $h_i \sigma_i$ in addition to the kernel, playing the role of a magnetic field), to preserve the $\mathbb Z_2$ symmetry of the problem (i.e. the symmetry of changing all the spins in a configuration $\sigma_i\to-\sigma_i$). Furthermore, expecting spins to be more correlated with neighboring ones, we simplify the model even more by choosing $W$ with a kernel of size $k$ small, such that the configuration of a given spin is modified only depending on the value of the spin itself and the configuration of the $k^2-1$ closest neighbors  (see Fig.\ \ref{fig_model}). We also impose the spatial symmetries of the system on the kernel, to respect invariance by $\frac\pi2$-rotations and reflections. 
Finally, as we focus on the Ising model with periodic boundary conditions, the system is translation invariant, and we use circular padding for the convolution kernel.
In total, our model with kernel size $k$ therefore consists of $(k+1)(k+3)/8$ independent parameters that will be optimized during the machine learning process.
In the following, we will consider only kernels of sizes $k=3$, $ 5 $, and $7$, corresponding to three, six, and ten free parameters. As we will see, the smallest kernel is already able to learn scale invariance and many properties of the renormalization group.

\subsection{Training}

The goal of the training procedure is to find the $U(\bs|\bmu)$ that minimizes the Kullback-Leibler divergence $D_\textrm{KL}$ between the true joint probability distribution $P(\bs,\bmu)$ and the downscaled-upscaled one $Q(\bs,\bmu)$, Eq.\ \eqref{eq:Q}, i.e.\ minimizing
\begin{equation}
    D_\textrm{KL}(P||Q) = \sum_{\bs,\bmu} P(\bs,\bmu) \ln\!\bigl[P(\bs,\bmu)/Q(\bs,\bmu)\bigr].
\end{equation}
This is equivalent to minimizing the cross-entropy $H(P,Q)=- \sum_{\bs,\bmu} P(\bs,\bmu) \ln\!\bigl[Q(\bs,\bmu)\bigr]$ and can be more conveniently rewritten as the minimization of the loss function
\begin{equation}
   \mathcal L(U)=- \sum_{\bs} P(\bs) \sum_{\bmu} D(\bmu|\bs)  \ln\!\bigl[U(\bs|\bmu)\bigr],
    \label{eq:loss}
\end{equation}
up to a constant independent of $U$. In practice, the average over all configurations is approximated by the Monte Carlo (MC) method, using a cluster algorithm to ensure correct sampling at criticality, in terms of $M$ configurations $\bs^{(a)}$, $a=1,\ldots,M$. The training configurations $\bs^{(a)}$ are of size $L_t^2$.\footnote{We used a standard Swendsen-Wang cluster algorithm \cite{Swendsen1987}, running at the critical temperature of the infinite system. We sampled Monte Carlo configurations every $L_t^{1/2}$ steps to ensure statistical independence of the configurations [the dynamical exponent of the algorithm for the two-dimensional (2D) Ising is smaller than $1/2$].} Assuming that $\bs^{(a)}$ corresponds to a given block-spin configuration $\bmu^{(a)}$ (of size $(L_t/2)^2$), we therefore have to optimize
\begin{equation}
    \mathcal L(U)\simeq -\frac{1}{M}\sum_{a=1}^M \ln\!\bigl[U(\bs^{(a)}|\bmu^{(a)})\bigr].
\end{equation}

If the block-spin procedure is not deterministic, as in the case of the majority rule used here, the sum over $\bmu$ in Eq.\ \eqref{eq:loss} is also sampled randomly, i.e.\ we choose one of the configuration $\bmu^{(a)}$ allowed from $\bs^{(a)}$ with probability $w_a=D(\bmu^{(a)}|\bs^{(a)})$. The loss function has to be weighted accordingly,
\begin{equation}
    \mathcal L(U)\simeq -\frac{1}{M}\sum_{a=1}^M w_a \ln\!\bigl[U(\bs^{(a)}|\bmu^{(a)})\bigr].
   \label{eq_loss}
\end{equation}
Other than this change of loss, the training procedure is the same. In the following, we have trained using the true loss (with non-trivial weight $w_a$) and a loss where we impose a uniform weight $w_a=1$ by hand.

\begin{figure}
    \centering
    \includegraphics[width=1\linewidth]{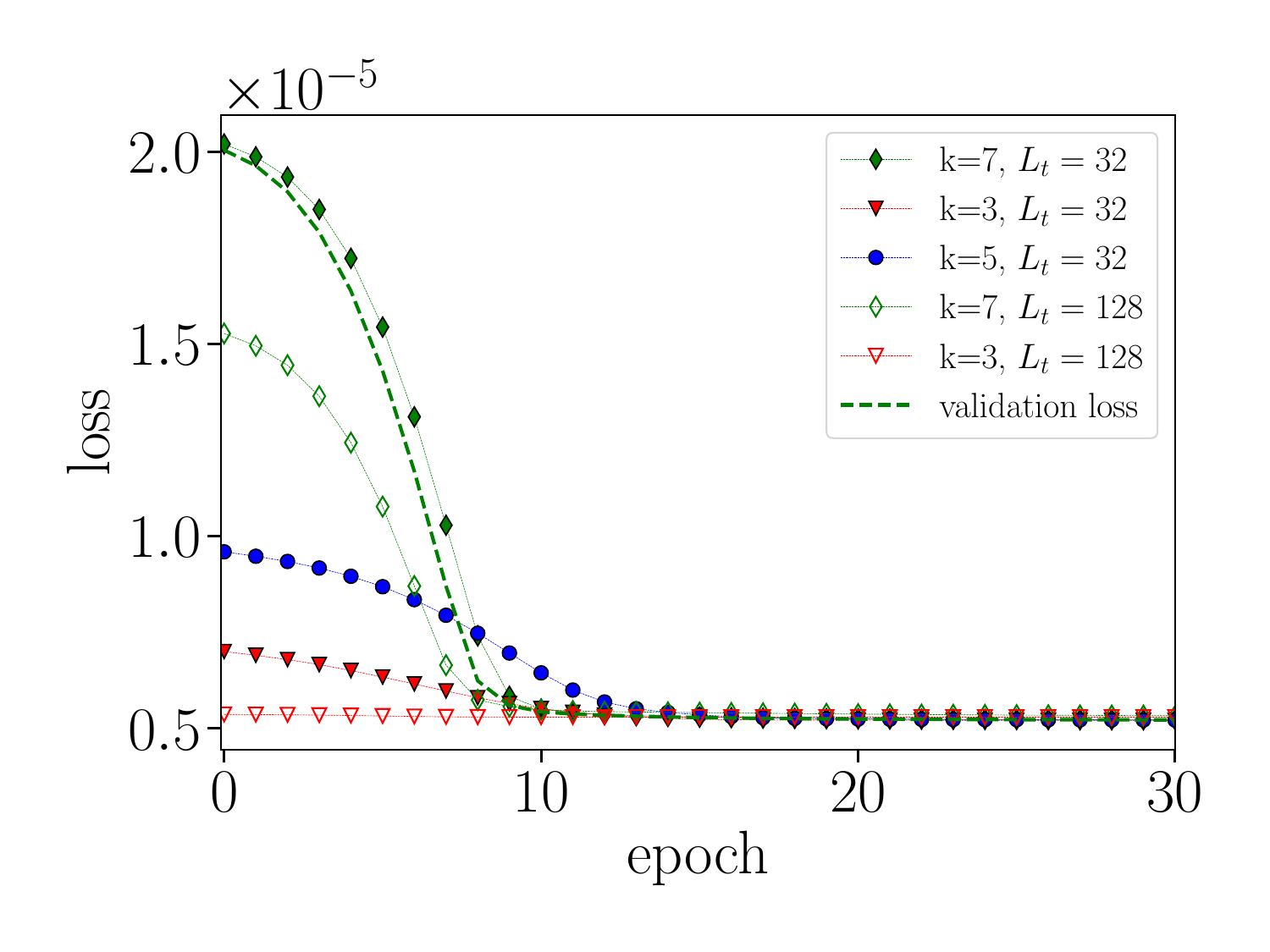}
    \caption{Training losses for different models. The learning rate was in all cases $0.01$, with $M=10^6$ and batch size $5\times 10^4$.
    The dashed line shows a typical comparison between the training and the validation loss, here for $k=7$, $L_t=128$, and $M=10^6$.  As we train the models we test the loss on the $\frac{1}{5} M$ configurations that are statistically uncorrelated with the training configurations. The validation loss closely follows the training loss, showing that the model is not overfitting.}
    \label{fig_losses}
\end{figure}

The optimization procedure is performed using the AdamW algorithm \cite{loshchilov2018decoupled}, with a fixed learning rate of $0.01$. The training minimizes the loss \eqref{eq_loss} of the generated spin configurations, averaged over $M$ Monte Carlo samples. Gradient updates are computed over mini-batches, and no learning rate decay or early stopping is used (see Fig.~\ref{fig_losses}).

\begin{table}
\begin{center}
\begin{tabular}{|c|c|}
\hline
\multicolumn{1}{|c|}{Hyperparameter} & \multicolumn{1}{c|}{Symbol} \\ \hline\hline
kernel size & $k$ \\ \hline
\shortstack[l]{linear size of training\\configurations} & $L_t$
\\ \hline
number of training configurations & $M$
\\ \hline
weighting during training & w.
\\ \hline
\shortstack[l]{linear size of the initial configuration\\before upscaling} & $L_0$
\\ \hline
number of upscaling & $n_\textrm{up}$
\\ \hline
\end{tabular}
\end{center}

\caption{Hyperparameters and their corresponding symbols. }
    \label{tab:symbols}
\end{table}

\subsection{Upscaling procedure}

We have trained our simple neural nets on the two-dimensional Ising model, Eq.~\eqref{eq:Ising}, at the critical point $K=\ln(1+\sqrt{2})/2$ \cite{Onsager44}. The goal is to see whether such a simple neural net can capture and reproduce scale invariance, both qualitatively and quantitatively.

We can use the nets described above to generate microscopic configurations $\bs$ by upscaling a given macroscopic configuration $\bmu$. Indeed, the probability of $\bs$ given $\bmu$ is just given by the learned $U(\bs|\bmu)$, and given $\bmu$ we can sample a plausible configuration $\bs$ by simply sampling $U(\bs|\bmu)$.

In previous works \cite{Efthymiou2019,Shiina2021,Bachtis2022}, the upscaling procedure was started from Monte-Carlo-sampled configurations $\bmu$ of finite size (for instance $L_0=16$).
More specifically, the starting configuration $\bmu$ was first sampled using a Monte Carlo simulation (say at criticality) and then the (complicated) neural nets were used to create new macroscopic configurations $\bs$ of twice the size $L_1=2L_0$. The procedure could then be repeated to obtain configurations of increasing sizes $L_n=2^n L_0$.

\begin{figure}[t!]
    \centering
    \includegraphics[width=1.0\linewidth]{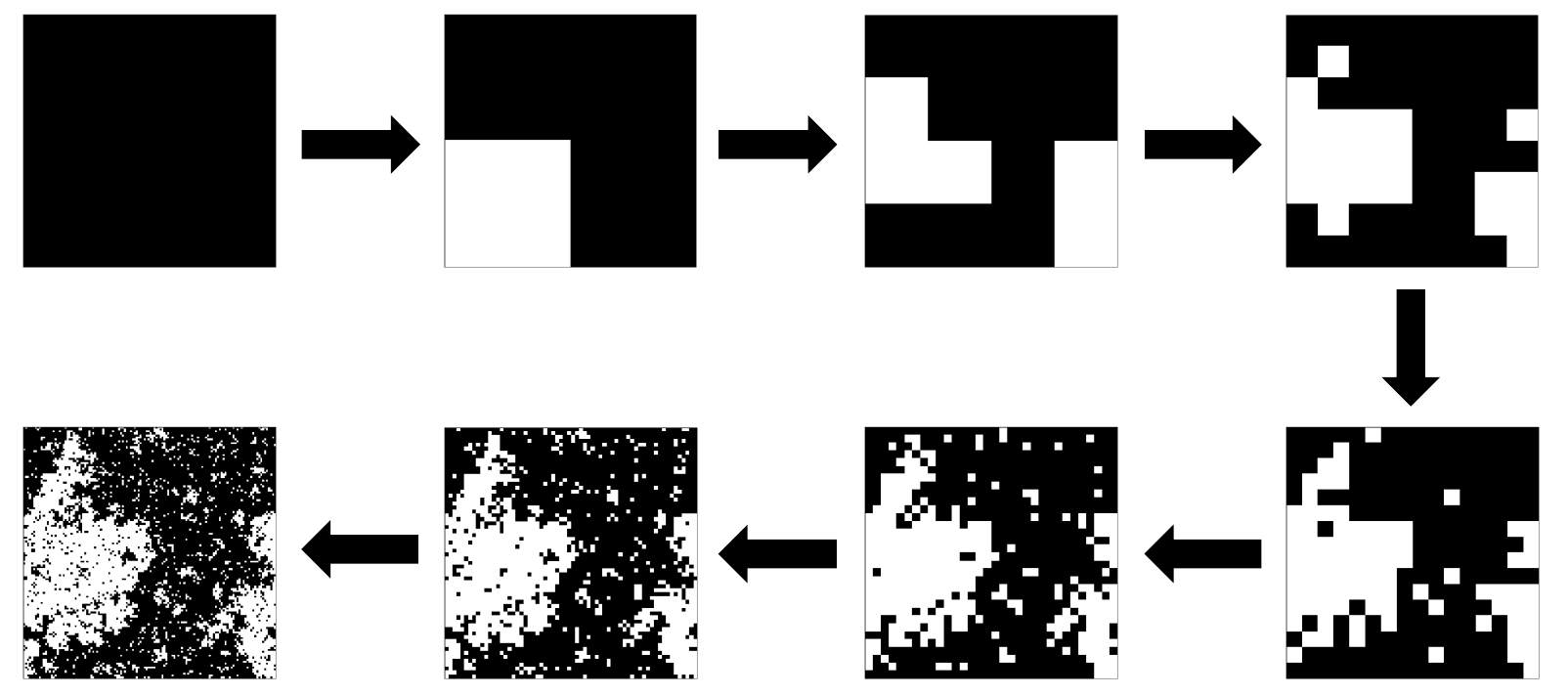}
    \caption{Example of the upscaling procedure. The starting point is a single spin ($L_0=1$), top left. Each upscaling is symbolized by an arrow, doubling the system's size.}
    \label{fig:example_upscaling}
\end{figure}

Here instead, if not stated otherwise, the starting point is a single spin (i.e. $L_0=1$) randomly chosen to be $\pm 1$ (with probability $\frac{1}{2}$). This single spin can then be upscaled $n_\textrm{up}$ times to obtain configurations on lattices of size $L=2^{n_\textrm{up}}$. We thus do not use the ``help'' of MC sampling to start from typical configurations at an intermediate size as previously done, but fully dream up configurations from one macroscopic block spin; see Fig.~\ref{fig_model} for the full procedure and Fig.~\ref{fig:example_upscaling} for an example of the generated configurations.
For practical reasons, we have restricted upscaling up to $n_\textrm{up}=7$ corresponding to lattices of size $L=128$. For all models nets, and all $L=2^{n_\textrm{up}}$, we generated $10^6$ configurations.

The list of all the hyperparameters used for training and producing new configurations is summarized in Tab.~\ref{tab:symbols}.

\section{\label{sec:FSS} Finite size scaling of global observables}

We start our analysis of upscaling with simple global observables that are typically used in numerical analysis of criticality (see, e.g., \cite{Landau_Binder_2014}). Indeed, the finite-size scaling of these quantities allows for devising the scaling properties of the system at criticality.

From the upscaled configurations $\bs$ of size $L$, defining the magnetization density $m=L^{-2}\sum_i\sigma_i$ and energy density $e=L^{-2}\sum_{\langle i,j\rangle} \sigma_i \sigma_j$,  we compute the magnetic susceptibility $\chi$, magnetization-energy cumulant $c_{me}$, heat capacity $c$, and the Binder ratio $U_4$.
\begin{equation}
    \begin{split}
        \label{eqs_susc}
    \chi &= L^2 ( \langle m^2 \rangle - \langle m \rangle^2 ), \\
    c_{me} &= L^2 \frac{( \langle |m||e| \rangle - \langle |m| \rangle  \langle |e| \rangle )}{\langle|m|\rangle}, \\
    c &= L^2 ( \langle e^2 \rangle - \langle e \rangle^2 ), \\
    U_4 &= 1-\frac{ \langle m^4 \rangle }{3 \langle m^2 \rangle ^2},
    \end{split}
\end{equation}
where  $ \langle \cdot \rangle $ is the average over the $10^6$ upscaled configurations. At criticality, for the two-dimensional Ising model, those quantities have the well-known exact finite-size scaling behaviors
\begin{equation}
    \begin{split}
    \chi &\propto  L^{\frac{\gamma}{\nu}},\\
    c_{me} & \propto  L^{\frac{1}{\nu}}, \\
    c &\propto \ln(L),
    \end{split}
\end{equation}
where  $\frac{\gamma}{\nu}=2-\eta=7/4$, $\frac{1}{\nu}=1$. Furthermore, $U_4$ tends to a universal constant the numerical estimate of which is $0.61069\ldots$ as $L\to \infty$  \cite{Nicolaides1988}.

\begin{figure*}
    \centering
    \includegraphics[width=1\linewidth]{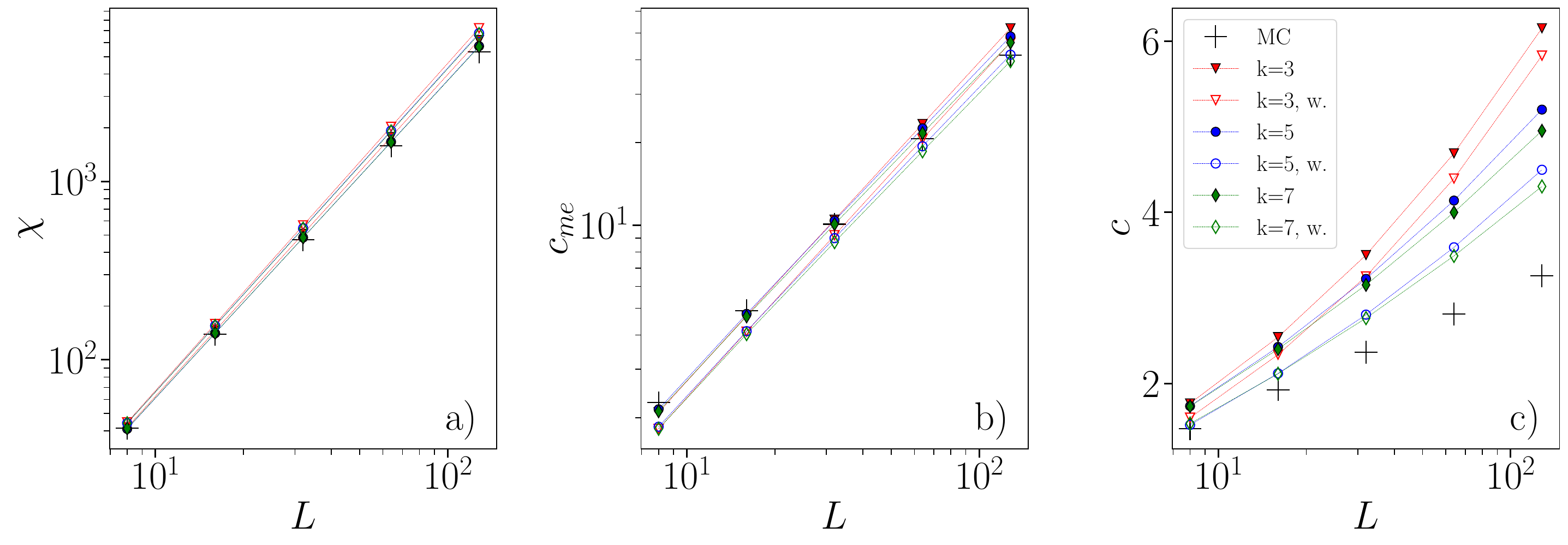}
    \caption{Susceptibilities obtained from various hyper-parameters as a function of the upscaled size: (a)  magnetic susceptibility $\chi$;  (b)  magnetization-energy cumulant $c_{me}$;  (c) heat capacity $c$. Plots in (a) and (b) are in log-log scale (where power law behavior displays as linear) while the plot in (c) is in log-normal scale (where the $\ln L$ dependence displays as linear). All models were trained on $M=10^6$ critical configurations of system size $L_t=32$.}
    \label{fig_susc}
\end{figure*}

Figure \ref{fig_susc} presents typical results for susceptibilities obtained with various model hyperparameter settings. The results are derived from upscaled data generated by single-layer models trained on $M = 10^6$ critical configurations of size $L_t = 32$. For comparison, we also include MC results obtained via the Swendsen-Wang cluster algorithm \cite{Swendsen1987}, obtained for the same system sizes (from $8$ to $128$) at criticality with the same statistics. This is sufficient statistics to ensure that the MC results are converged. 

In general, there is good agreement between the upscaled and MC data across most quantities. Specifically, both the susceptibility, $\chi$, and the magnetization-energy cumulant, $c_{me}$, exhibit clear power-law scaling, appearing as straight lines in the log-log plot. As could have been expected, increasing the convolution kernel size, $k$, results in systematic improvements, with model results converging more closely to the MC data. This effect is particularly pronounced for the heat capacity, which shows notable gains in accuracy with larger $k$ values.
However, accurately reproducing the logarithmic scaling behavior of the heat capacity proves more challenging. The upscaled data exhibits weak power-law behavior, which deviates slightly from the expected straight line in the log-normal plot, with the slope diminishing as $k$ increases.

\begin{center}
\begin{table}[t!]
\begin{tabular}{|c |c | c | c |} 
 \hline
 & $\gamma/\nu$  $(32,8)$ & $\gamma/\nu$ $(64,16)$ & $\gamma/\nu$ $(128,32)$ \\
 \hline\hline
 MC & 1.754 & 1.751 & 1.749 \\
 \hline\hline
 k=3 &1.803 & 1.804 & 1.805 \\ 
 \hline
 k=5 &1.781 & 1.782 & 1.782\\
 \hline
 k=7 &  1.777 & 1.778 & 1.778 \\
 \hline
 k=3, w. & 1.834 & 1.835 & 1.836 \\
 \hline
 k=5, w. & 1.815 & 1.816 & 1.816  \\ 
 \hline
 k=7, w. & 1.812 & 1.812 & 1.812 \\ 
 \hline
 \end{tabular}
 \vspace{.3cm}
 
 \begin{tabular}{|c |c | c | c |} 
\hline
& 1/$\nu$, $(32,8)$ & 1/$\nu$ $(64,16)$ & 1/$\nu$ $(128,32)$ \\
\hline
\hline
MC & 1.077 & 1.038 & 1.020 \\
\hline\hline
k=3 & 1.163 & 1.157 & 1.155 \\
\hline
k=5 & 1.138 & 1.120 & 1.112 \\
\hline
k=7 & 1.129 & 1.106 & 1.096 \\
\hline
k=3, w. & 1.169 & 1.164 & 1.163 \\
\hline
k=5, w. & 1.138 & 1.116 & 1.107 \\
\hline
k=7, w. & 1.126 & 1.103 & 1.093 \\
\hline
\end{tabular}
 \caption{Exponents $\gamma/\nu$ and $1/\nu$ determined from $\chi$ and $c_{me}$ using the two values $(L,L/4)$ shown in parenthesis.  The last digit displayed is reliable, given the statistics. The exact exponents are $\gamma/\nu=1.75$ and $1/\nu=1$. }
\label{tab_exp}
\end{table}
\end{center}

To quantitatively assess scaling behavior, we present in Table~\ref{tab_exp} the values of the critical exponents $\gamma/\nu$ and $1/\nu$, obtained by fitting data from lattices of size $L/4$ and $L)$. These results from upscaled configurations are compared to those obtained from MC simulations. Increasing the kernel size $k$ further enhances accuracy, with larger kernels yielding results closer to the exact exponents.

\begin{figure}
    \centering
    \includegraphics[width=1\linewidth]{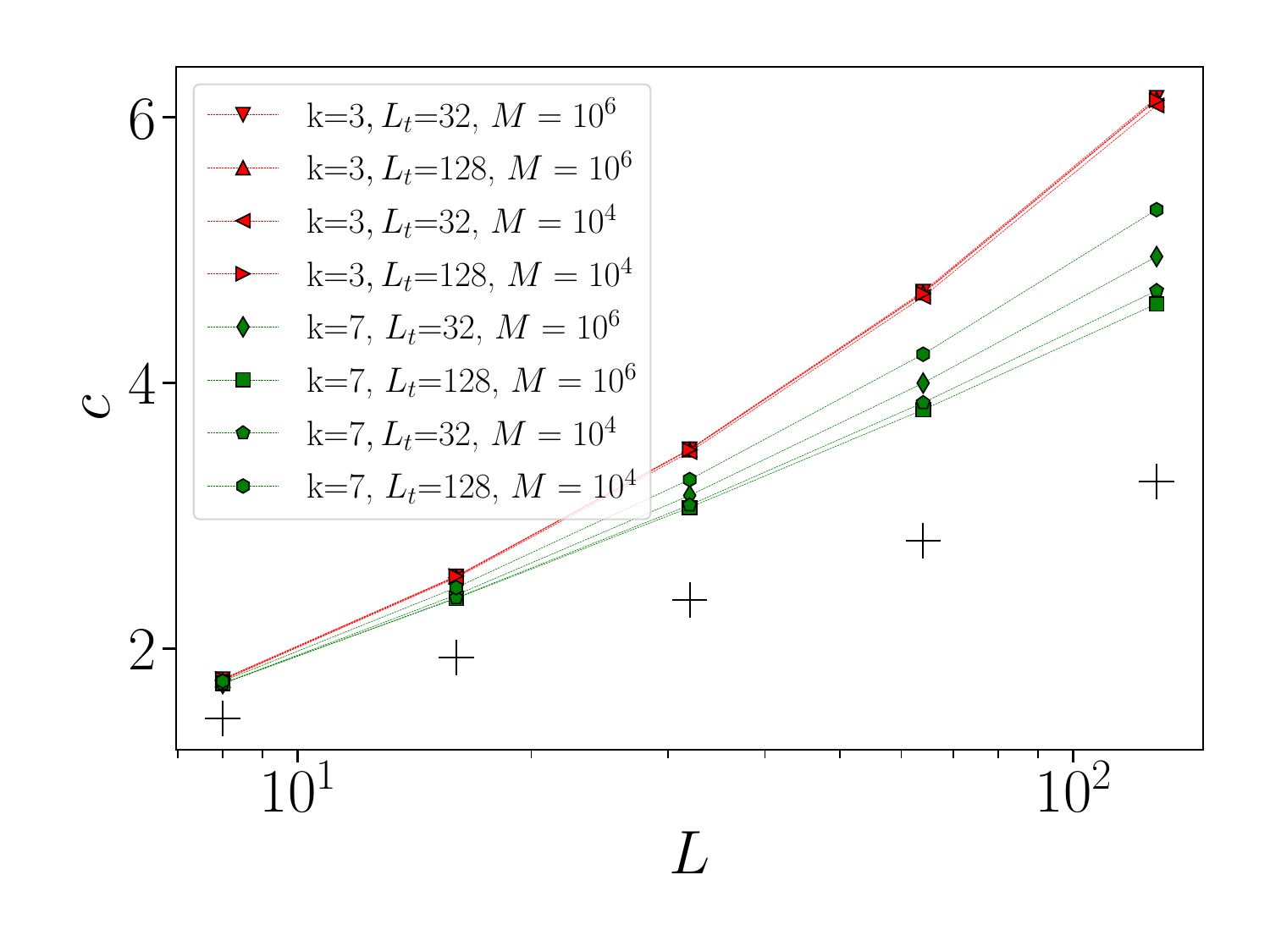}
    \caption{
    Dependence of the heat capacity $c$ calculated from the upscaled configurations on the size of the training configurations ($L_t$) and the statistics of the training set ($M$). Black crosses correspond to MC data.}
    \label{fig_hyperp}
\end{figure}

While the results show promising convergence with the exact universal values, we observe some scatter across models with different hyperparameters, indicating potential for further refinement. To address this, we examined the influence of kernel size ($k$), training lattice size ($L_t$), and training set size ($M$) on the accuracy of estimated exponents. As shown in Fig.~\ref{fig_hyperp}, which uses heat capacity as a representative metric, these factors have distinct effects.

For models with a kernel size of $k=3$, increasing $L_t$ or $M$ has minimal impact on scaling accuracy, likely due to the limited number of free parameters constraining the network’s learning potential. In contrast, models with a larger kernel size of $k=7$ demonstrate greater adaptability. Increasing $M$ significantly improves accuracy, suggesting the network benefits from a larger training set. However, increasing $L_t$ beyond 32 does not yield systematic improvement; even at $L_t = 128$, variations in model architecture lead to more significant deviations than changes in lattice size. This suggests that beyond a certain lattice size, kernel size, and training set size are more influential in refining the accuracy of the critical exponents.

\begin{figure}[t]
    \centering
    \includegraphics[width=1\linewidth]{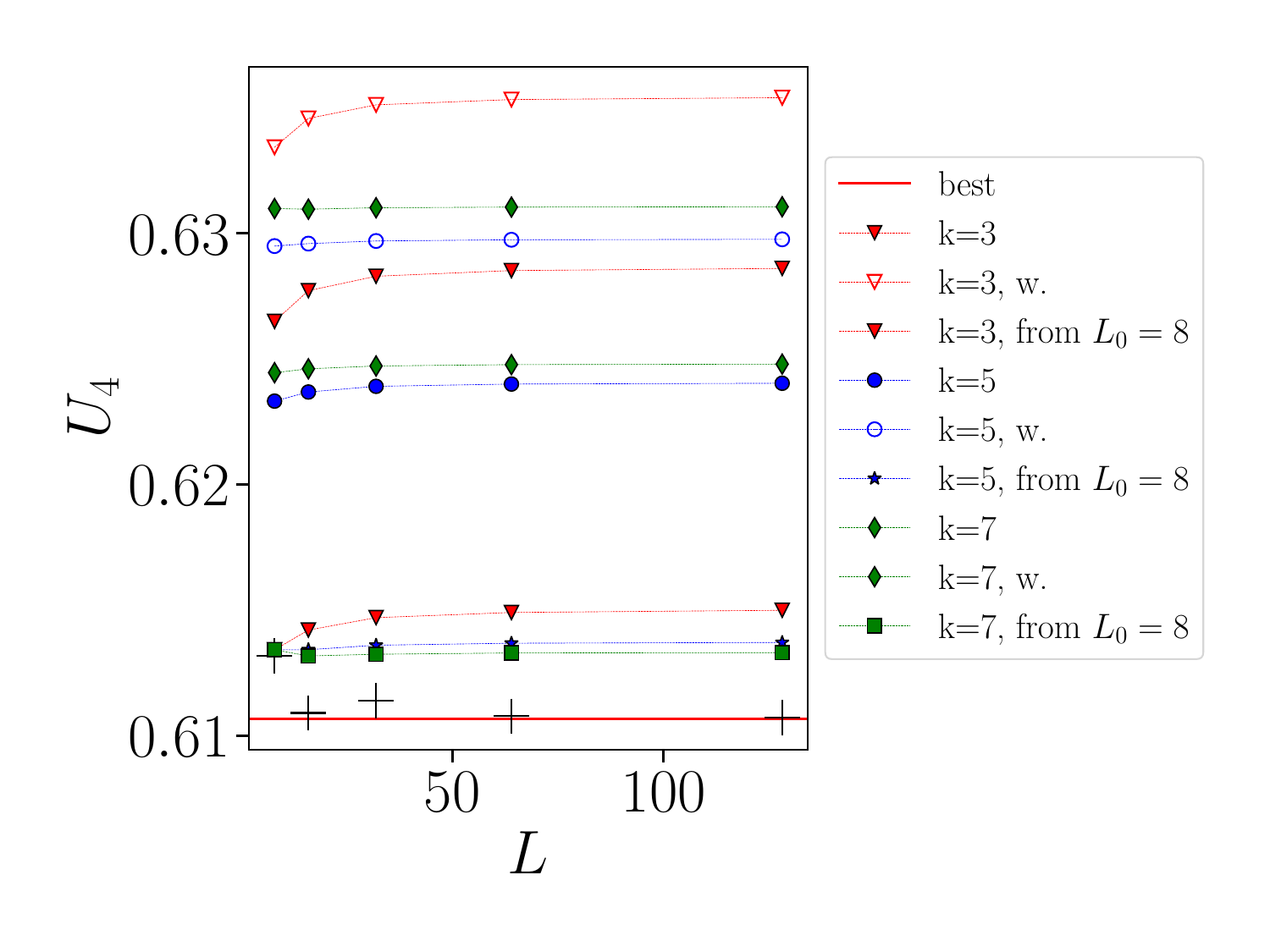}
    \caption{
    Binder ratio computed from upscaled configurations and compared with the best-known value \cite{Nicolaides1988} and our Monte Carlo data (black crosses). }
    \label{fig:binder}
\end{figure}

The results for the Binder ratio are comparable to those observed for the critical exponents of the susceptibilities, see Fig.~\ref{fig:binder}. The scatter from the established reference value varies among the models, with a discrepancy of a few percent. To assess the reason of this discrepancy, we have also upscale data from $10^6$ independent MC configurations from simulations of a system of size $L_0=8$ in the spirit of \cite{Efthymiou2019,Shiina2021,Bachtis2022}, instead of a system of size $L_0=1$.
This noticeably improves the Binder ratio to an error of less than a percent. However, upscaling from $L_0=8$ does not consistently lead to better determinations of the critical exponents from susceptibilities.
Note that the value of $U_4$ from the upscaled data from $L_0=8$ MC configurations is very close to the exact value of the Binder ratio at $L_0=8$, and almost independent of the number of upscaling. That is, it is as if the upscaling procedure does not change the value of the Binder ratio of the MC data, while still obeying scale invariance (since $U_4$ is a ratio of correlation functions that scale). 

\begin{figure}
    \centering
    \includegraphics[width=1\linewidth]{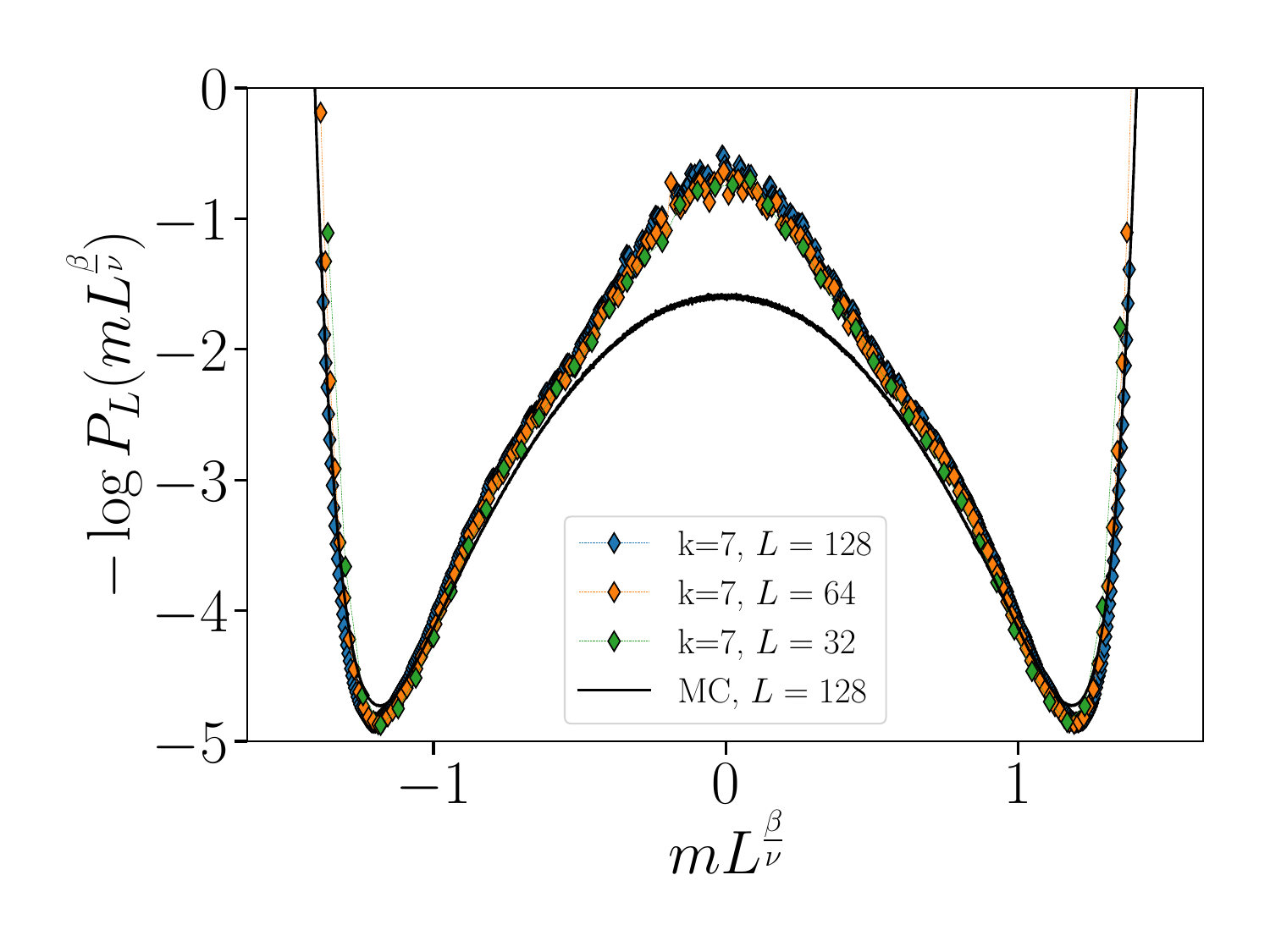}
    \caption{Log of the probability distribution of the order parameter as a function of $m L^{\beta/\nu}$, obtained from MC simulations of lattice size $L=128$, and upscaled configurations (upscaled to $L=32$, $L=64$ and $L=128$)  (hyperparameters: $k=7$, $L_t=32$, and $M=10^6$). The probability is scale-invariant, as seen from the collapse of the three curves onto a single one. We use the exact exponent $\beta/\nu=1/8$ for rescaling all the data. }
    \label{fig_I_upL}
\end{figure}

\begin{figure}
    \centering
    \includegraphics[width=1\linewidth]{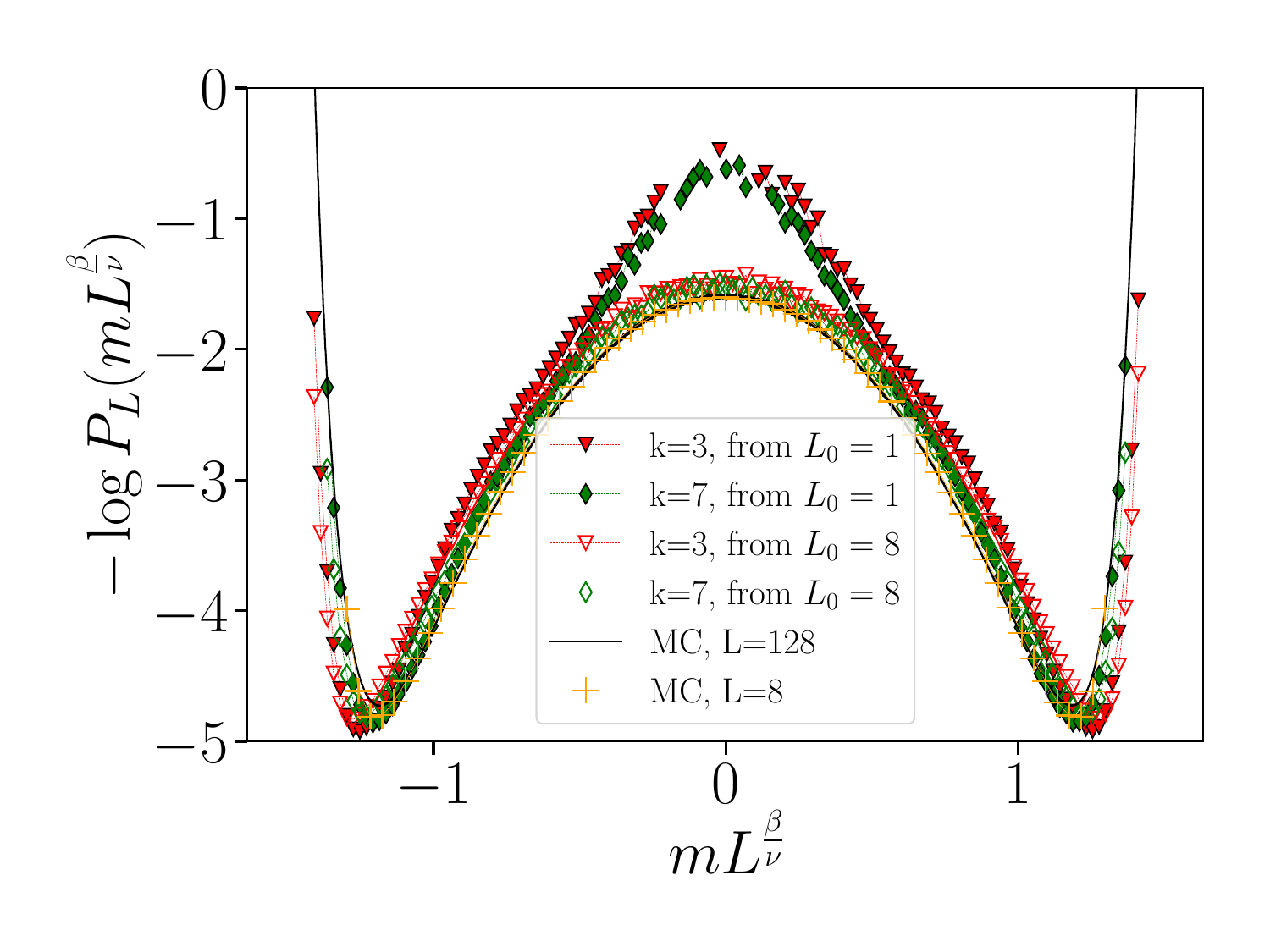}
    \caption{
    Log of the probability distribution of the order parameter for a system of size $L=128$, obtained from MC and upscale data. In the latter case, upscaling from $L_0=8$ instead of $L_0=1$ improves the results, whereas the kernel size is rather irrelevant. Models trained with $L_t=32$ and $M=10^6$. We use the exact exponent $\beta/\nu=1/8$ for rescaling all the data.
     }
    \label{fig_I}
\end{figure}

This can be verified by computing the probability distribution (PDF) of the order parameter $P_L(m)$. The true PDF obeys at criticality the scaling law (for sufficiently large $L$)
\begin{equation}
    P_L(m)=L^{\beta/\nu}\tilde P(L^{\beta/\nu} m),
\end{equation}
with $\beta/\nu=1/8$ for Ising 2D, and $\tilde P$ is a universal function that has been studied thoroughly both numerically and theoretically \cite{Binder1981a,Bouchaud1990,Bruce1992,Chen1996,Tsypin2000,Xu2020, balog_critical_2022}. Note that the $P_L(m)$ obtained from MC simulations at $L=128$ is converged up to better than $1\%$ to the thermodynamic limit.
 Fig.~\ref{fig_I_upL} shows the logarithm of the PDF, $\ln P_L(m)/P_L(0)$ obtained from upscaling system of size $L_0=1$. Remarkably, we see that it obeys the scaling law very well and is completely size-independent (i.e.\ independent of the number of upscaling). This explains why the correlation functions of the magnetization (such as $\chi$) also obey their scaling laws very well. However, we also see that the shape of the PDF is slightly off (in particular in the middle), which explains why the Binder ratio is off. Nevertheless, the collapse across system sizes shows that the networks do capture the RG-relevant structure of the fixed point, whereas the remaining discrepancies arise from the extreme upscaling path $1\!\to\!128$. This is confirmed by upsampling from MC configurations with $L_0=8$, see Fig.~\ref{fig_I}.  We observe that the shape of the so-obtained PDF is much improved. This can be understood by remarking that the PDF has quite weak finite-size corrections, meaning that the PDF at $L_0=8$ from which the network upscale is already quite similar to that of a system of size $L=128$. Therefore, the network, which as we have shown is good at generating scale invariant quantities, keeps the same global shape, with the proper scaling of the magnetization.

\section{\label{sec:MCRG} Real space renormalization group analysis of upscaled configurations}

We have observed that susceptibilities obtained from the upscaled data yield highly stable results, effectively capturing global properties during the upscaling process. To subject the upscaled configurations to a more stringent evaluation, we conduct an RSRG analysis on them \cite{Wallace1974, Peretto1976,Swendsen1979, Swendsen1984, Baillie1992, Brandt2001, Ron2002, Ron2005, Ron2017}. It is important to note that our objective is not to improve the RSRG block-spin transformation using machine learning, as previously done in \cite{Chung2021}. Instead, we employ RSRG solely as a tool to scrutinize and analyze the properties of the upscaled configurations. In particular, we will use the same majority rule used throughout the manuscript to coarse-grain.

To analyze whether the neural network has learned the renormalization group properties associated with downscaling, we take inspiration from Monte Carlo RG (MCRG), which is a way to approximate the (exact) RSRG transformation a la Kadanoff. Starting from the RSRG transformation defined in Sec.~\ref{sec:upscaling}, $\K_{n+1}=\RR(\K_n)$ (where the index $n$ denotes the number of blocking or the downscale step), with fixed point $\K_\star$ such that $\K_\star=\RR(\K_\star)$, the scaling properties of the critical system can be obtained by analyzing the linearization of the RSRG transformation close to the fixed point,
\begin{equation}
    K_{\alpha,n+1}-K_{\alpha,\star}=\sum_{\beta}\mathcal{T}^n_{\alpha,\beta}(K_{n,\beta}-K_{\beta,\star}).
\end{equation}
The matrix $\mathcal{T}$ gives us information about the flows near criticality and hence about the critical exponents. Diagonalizing it for $\K_n\simeq \K_\star$, the two largest eigenvalues $e_1$ and $e_2$ are related to the exponents $y_h=\frac{d+2-\eta}{2}$ and $y_{\tau}=\frac{1}{\nu}$ by $y_h=\frac{\ln e_1}{\ln 2}$ and $y_{\tau}=\frac{\ln e_2}{\ln 2}$, where $2$ is the downscaling factor. Here the $y_h$ exponent is related with an eigenvector which is odd and the exponent $y_{\tau}$ with an eigenvector which is even with respect to reflections of spin degrees of freedom.

The matrix $\mathcal{T}^n_{\alpha,\beta}$ can be determined from the correlation functions between different couplings since \cite{Swendsen1979}
\begin{equation}
   \mathcal{T}^n_{\alpha,\beta}=\frac{\partial K_{\alpha,n+1}}{\partial K_{\beta,n}}=\sum_{\gamma} \frac{\partial K_{\alpha,n+1}}{\partial \langle S_{\gamma}(\bmu)\rangle} \frac{\partial \langle S_{\gamma}(\bmu)\rangle}{\partial K_{\beta,n}}.
\end{equation}
Here $S_{\gamma}(\bmu)$ is a spin operator in terms of a coarse-grained configuration $\bmu$ obtained from a finer-grained configuration $\bs$, and the average is done over the probability distribution of the latter. 
Using
\begin{eqnarray}
\mathcal{M}^n_{\gamma,\beta} &=& \frac{\partial \langle S^n_{\gamma}(\bmu)\rangle}{\partial K_{\beta,n}}\\
&=&\langle S^n_{\gamma}(\bmu)S^n_{\beta}(\bs) \rangle-\langle S^n_{\gamma}(\bmu)\rangle \langle S^n_{\beta}(\bs) \rangle\\
\mathcal{N}^{n+1}_{\alpha,\gamma} &=& \frac{\partial \langle S^n_{\gamma}(\bmu)\rangle}{\partial K_{\alpha,n+1}}\\
&=& \langle S^{n+1}_{\alpha}(\bmu)S^n_{\gamma}(\bmu) \rangle-\langle S^{n+1}_{\alpha}(\bmu)\rangle \langle S^n_{\gamma}(\bmu) \rangle,
\end{eqnarray}
we then have $\mathcal{T}=\mathcal{M}^{-1}\mathcal{N}$. We immediately see that calculating this quantity entails calculating many different susceptibility-like quantities. Hence an analysis of the upscaled configurations using the matrix $\mathcal{T}$ constitutes a much more stringent test than simply calculating the susceptibilities of Eq.~\eqref{eqs_susc}.

The advantage of this formulation is that the matrix $\mathcal{T}$ can be obtained from correlation functions (instead of an abstract transformation in coupling space). In practical calculations, three main factors limit the precision of the method: (a) We are limited to calculation on finite lattices, (b) we need to choose the number of couplings to take into consideration, and (c) the number of configurations used to approximately compute the correlation functions is finite.

Despite these limitations, the RSRG works extremely well on the 2D Ising model for the determination of the critical exponents. As a benchmark, we have used the RSRG on lattices from $L=32$ to $L=1024$ for the 2D critical Ising model, with $10^6$ MC configurations for each case. We have restricted the space of couplings to $36$, i.e.\ all couplings four spins that fit onto a $3\times 3$ plaquette (see, e.g., \cite{Baillie1992}). 
The results for the three most important universal exponents are given in Table \ref{tab_MCRGexp}. If the $\mathcal{M}$ matrix was computed from couplings determined at the $n$th downscale step then the $\mathcal{N}$ matrix was computed from the couplings determined at $n$th step and from the $(n+1)$th step. We find that even at $n=1$, the couplings are already very close to the fixed point, i.e.\ $\K_1\simeq \K_\star$.  
Furthermore, the exponents obtained from MCRG are very close to the exact ones, and in good agreement with that obtained from finite size scaling (Tab.~\ref{tab_exp}).

\begin{center}
\begin{table}
\begin{tabular}{|c|c|c|c|}
  \hline
   $L$ & $y_h$ (exact: 1.875) & $1/\nu$ (exact: 1) & $\omega$ (exact: -1) \\
  \hline
  32 & 1.87537(5) & 0.931(2) & $-$1.35(5) \\
  128 & 1.87582(4) & 0.973(3) & $-$1.08(5) \\
  512 & 1.87599(4) & 0.992(2) & $-$1.04(4) \\
  1024 & 1.87600(4) & 0.997(2) & $-$1.03(4) \\
  \hline
\end{tabular}
\caption{Exponents $y_h$, $y_{\tau}=\frac{1}{\nu}$, and $\omega$ determined from RSRG on lattices of various sizes using $10^6$ MC configurations. The first two eigenvalues are easily identified as they are most relevant (as well as using their symmetry under spin flips). The leading correction $\omega$ can be identified by symmetry, however, it is higher in the spectrum, because several eigenvalues corresponding to redundant operators precede it. The number in parentheses is the error bar on the last digit, given the statistics of $10^6$ MC configurations.}
\label{tab_MCRGexp}
\end{table}
\end{center}

\begin{figure*}
    \centering
    \includegraphics[width=1.0\linewidth]{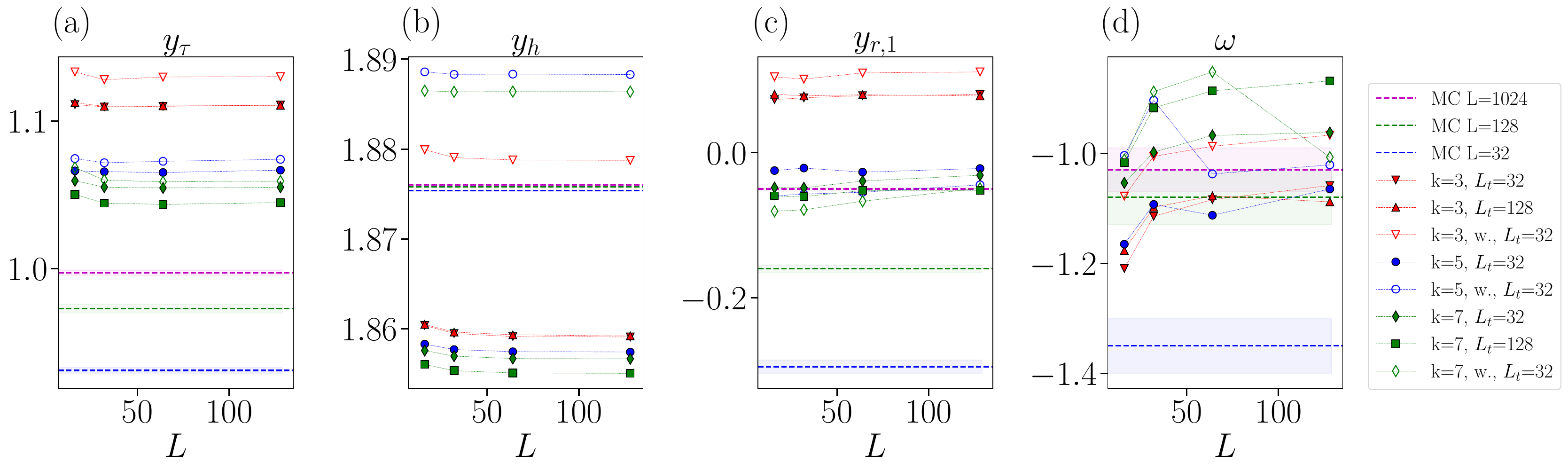}
    \caption{
    Exponents obtained from the eigenvalues of the RSRG matrix $\mathcal{T}$ from the configurations upscaled from $L=1$ to $L=128$ and calculated on $10^6$ configurations: (a) $y_{\tau}=1/\nu$, (b) $y_h=\frac{d+2-\eta}{2}$, (c) most relevant redundant eigenvalue, and (d) the correction-to-scaling exponent $\omega$. The shaded areas around the MC results give the statistical fluctuations, given the statistics of $10^6$ MC configurations. }
    \label{fig_up_4egv}
\end{figure*}

We now apply the same procedure to our models. For this, we upscale $n_\textrm{up}$ times (starting from $L_0=1$) to compute the matrix $\mathcal M$, and then downscale once to compute the matrix $\mathcal{N}$, and then the matrix $\mathcal{T}$,  using $10^6$ configurations of upscaled configurations in total.
The $\mathcal{T}$-matrix eigenvalues are shown in Figs.~\ref{fig_up_4egv}(a) to \ref{fig_up_4egv}(d). 
The stability of the critical exponents obtained from finite-size scaling to upscaling is corroborated by this independent method. However, a slight scatter is observed, particularly compared to the RSRG from MC configurations obtained for similar lattice sizes. Notably, increasing the kernel size or the training system size does appear improve the results. This is summarized in Table~\ref{tab_FSS_MCRG}, where we also recall the results obtained from finite-size scaling, written in terms of the exponents $\eta$ and $\nu$. In particular, the exponents determined from RSRG are systematically different from those obtained by finite-size scaling (by a few percent for $\nu$ and $20-30\%$ for $\eta$). Whether increasing kernel sizes even more would resolve this tension remains an open question.

\begin{center}
\begin{table}[h!]
\begin{tabular}{|c |c | c | c | c |} 
 \hline
  & $\eta$-FSS & $\eta$-RSRG & $1/\nu$-FSS & $1/\nu$-RSRG \\
 \hline\hline
 MC &   0.251  & 0.248 & 1.020 & 0.973 \\
 \hline\hline
 k=3 & 0.195 & 0.282 & 1.155  & 1.111\\ 
 \hline
 k=5 & 0.218 & 0.285 &  1.112 & 1.067\\
 \hline
 k=7 & 0.222 & 0.287 & 1.096 & 1.055\\
 \hline
 k=3, w. & 0.164 & 0.243 & 1.163  & 1.130 \\
 \hline
 k=5, w. & 0.184 & 0.223 & 1.107 & 1.074\\ 
 \hline
 k=7, w. & 0.188 & 0.227 & 1.093  & 1.059\\ 
 \hline
\end{tabular}
 \caption{Critical exponents $\eta$ and $\nu$ obtained from finite-size scaling (FSS) and the RSRG. The FSS results are from the last column of Table~\ref{tab_exp}. The RSRG results are obtained from upscaled configurations to size $L=128$. The last digit displayed is reliable, given the statistics.}
\label{tab_FSS_MCRG}
\end{table}
\end{center}

Interestingly, our upscaling procedure also reconstructs the higher eigenvalues of the $\mathcal{T}$ matrix remarkably well. We give two examples that we see as relevant.

Fig.~\ref{fig_up_4egv}(c) shows the exponent $y_{r,1}=\frac{\ln e_3}{\ln 2}$ derived from the third eigenvalue of the matrix $\mathcal{T}$. We associate this eigenvalue with the most relevant redundant eigenvalue. Redundant eigenvectors and eigenvalues are non universal and related to the specifics of the renormalization procedure, such as the specific way in which we do downscale in the particular case.
We have not been able to find the determination of this exponent in the literature. However, from our calculations, it seems to vanish as we increase the system size. 
Also note that only the $k=3$ results group near $0.1$ while all larger kernels seem to group in the vicinity of zero. 

Finally, Fig.~\ref{fig_up_4egv}(d) shows the leading correction-to-scaling exponent $\omega$, the exact value of which is $-1$. This exponent is determined from the sixth eigenvalue of the matrix $\mathcal{T}$, which in our calculation is related to the second most relevant eigenvector in the spectrum that is even under reflections of spin degrees of freedom. The determination of this eigenvalue from MC configurations becomes visibly affected by the statistics. This is also true for the upscaled data, as its value is not stable to upscaling. However, we see that the scatter is smaller than the difference between the exact value and the MC estimate at small size ($L=32$).

\section{\label{sec:complicated} Impact of architectural complexity on upscaling}

Our results demonstrate that minimal, single-layer convolutional networks can successfully generate scale-invariant configurations that reproduce the universal features of critical phenomena. This naturally raises the following question: Does increasing model complexity improve performance?

To address this, we considered more elaborate network architectures, specifically a two-layer, 32-channel local convolutional network similar in spirit to those used in prior work \cite{Efthymiou2019,Shiina2021,Bachtis2022}. The architecture is as follows: (i) nearest neighbor upscaling from $L/2$ to $L$ as used in the single-layer models, (ii) one convolutional layer of kernel size $k$ from 1-to-32 channels of stride one and circular padding,\footnote{Here the convolutional kernels are not symmetrized to implement the geometric symmetries of the lattice.} (iii) a nonlinearity implemented with a rectified linear unit, and (iv) one convolutional layer of kernel size $k$ from 32 to 1 channel of stride one and circular padding. The upscaled configurations are then sampled using a sigmoid, similar to Eq.~\eqref{eq:Psm}.

\begin{figure}
    \centering
    \includegraphics[width=1\linewidth]{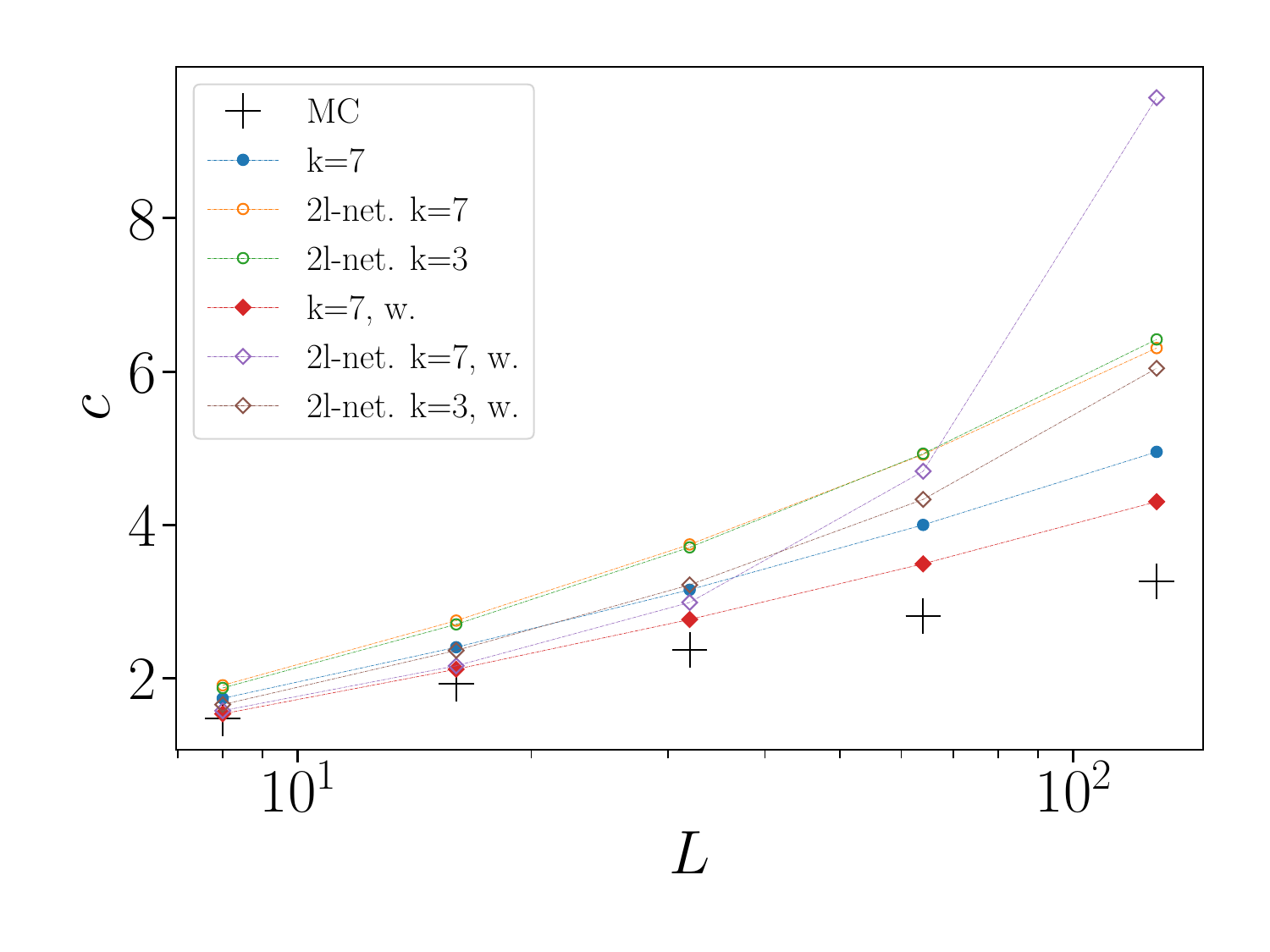}
    \caption{Heat capacity from the upscaled configurations from the 2-layer model with kernel size $k=3$ and $7$, and with or without weight in the training. The results from the simpler model with the same kernel sizes are also shown for comparison. For all data $L_t=32$ and $M=10^6$.
    }
    \label{fig:2layer}
\end{figure}

When trained to similar loss values, the two-layer network performs comparably to the single-layer models on global observables such as susceptibility and Binder ratio. However, we find that it consistently underperforms on quantities involving energy fluctuations, such as the heat capacity and magnetization-energy cumulant (see Fig.~\ref{fig:2layer}). This underperformance persists even when increasing the training set size or kernel width. While the root cause of this limitation is not yet fully understood, it suggests that the added depth and number of parameters do not necessarily improve the model’s ability to reproduce fine thermodynamic structure.

Moreover, the expected benefits of overparametrization (such as improved generalization or better modeling of long-range correlations) are not observed. The network appears to behave similarly to small-kernel single-layer models in terms of both scaling exponents and probability distributions. This may indicate that the architecture's representational capacity is not effectively exploited during training or that the essential features of criticality are already captured by local operations with minimal structure. Furthermore, the complexity of such architecture hinders the interpretability of the networks.

Our finding that increased network depth or complexity does not improve critical-point generation is consistent with observations in other contexts: Simpler or symmetry-constrained models often capture the essential physics more effectively \cite{Casert2019}. Highly parametrized networks may learn spurious, non-universal features, whereas minimalist high-compressing models or those explicitly encoding symmetries tend to focus on the relevant collective variables. Indeed, recent interpretability studies of phase-classifying neural networks reveal that such models internally learn representations of conventional order parameters \cite{Casert2019,Alamino2024}, reinforcing the idea that the crucial information governing a phase transition is remarkably low dimensional. In our case, a three-parameter filter is sufficient to discover magnetization and critical fluctuations, much like Kadanoff's block-spin construction which distills large-scale order from local degrees of freedom. 

While more advanced architectures (e.g.\ graph-based neural networks or attention mechanisms) could be applied to this problem, our results indicate that their added complexity is unnecessary for the 2D Ising critical point \cite{Ma2024,Kara2021,Chang2023}. Incorporating known physical symmetries and locality constraints directly into a simple model appears to improve performance over a brute-force increase in model size.

This underscores a central message: Complexity is not required to achieve scale-invariant generative modeling of critical systems. In fact, the essence of universality can be learned and generatively reproduced by surprisingly simple models, which offers advantages in interpretability, stability, and potential applicability to other systems.

\section{Conclusion \label{sec:conclusion}}

We have shown that simple, local neural networks can effectively learn to invert the renormalization group coarse-graining procedure in the two-dimensional Ising model, generating scale-invariant spin configurations from minimal input. Strikingly, networks with as few as three parameters are sufficient to reconstruct critical observables and reasonable scaling exponents. Through both finite-size scaling and real-space renormalization group analysis, we demonstrated that the generated ensembles preserve key features of universality, including nontrivial RG eigenvalue spectra and very good probability distributions of the order parameter. 

Perhaps most unexpectedly, increasing model complexity via additional convolutional layers and non-linearities does not improve performance and can even hinder the faithful reconstruction of critical statistics. This observation suggests that the essence of universality is encoded in simple, local, and symmetric transformation rules, much like those generating fractal structures. In this sense, our neural networks act as generative models for criticality, discovering scale-invariant ensembles from minimal assumptions and limited microscopic data.

These results open several directions for future research. From a theoretical point of view, they raise the question of what it truly means to ``invert'' the RG flow: Can such minimal models be understood analytically, or classified according to the universality class they encode? From a practical perspective, they suggest new ways to simulate critical systems, i.e., without Hamiltonians or explicit dynamics, by learning the statistical structure of scale invariance directly.  Extending this approach to higher dimensions, disordered systems, or systems out of equilibrium could shed new light on the generality of this framework and its analytical properties. In particular, introducing a handful of additional parameters into the model and training it to upscale at arbitrary temperatures could be an exciting avenue to understand how scale invariance is encoded by the models. 
Ultimately, while our tests on the generated upscaled configurations show that they are not sampled from the exact critical probability distribution $P_\star$, it remains an open question whether alternative machine-learning architectures could generate configurations genuinely indistinguishable from Monte Carlo ones. Achieving this would amount to capturing the full statistical content of criticality, and would provide a compelling demonstration that such models can faithfully reproduce all aspects of critical behavior.

\begin{acknowledgments}

 A.R. and I.B. thank Giulio Biroli for valuable discussions. T.I.\ and I.B.\ thank Domagoj Vlah for valuable discussions and for generously making computational resources available. A.R. and I.B. were supported by Project No.\ HRZZ-IP-10-2022-9423. I.B. acknowledges support from the project ``Implementation of
cutting-edge research and its application as part of the Scientific Center of Excellence for Quantum and Complex
Systems, and Representations of Lie Algebras'', Grant No. PK.1.1.10.0004, co-financed by the European Union
through the European Regional Development Fund - Competitiveness and Cohesion Programme 2021-2027. A.R. has benefited from the financial support of the Grant No. ANR-24-CE30-6695 FUSIoN. I.B. and T.I. wish to acknowledge the support of the INFaR and FrustKor projects financed by the EU through the National Recovery and Resilience Plan (NRRP) 2021-2026 and the KaCIF project, co-financed by the Croatian Government and the European Union through the European Regional Development Fund---Competitiveness and Cohesion Operational Program (Grant No. KK.01.1.1.02.0012). 
\end{acknowledgments}

\bibliography{bibli}
\bibliographystyle{apsrev4-2} 

\end{document}